\documentclass[letterpaper, 10 pt]{ieeeconf}  

\IEEEoverridecommandlockouts                              

\usepackage[utf8]{inputenc}
\usepackage[style=ieee,isbn=false,url=false,eprint=false]{biblatex}

\usepackage{amsmath,amssymb,amsfonts}
\usepackage{algorithmic}
\usepackage{graphicx}
\usepackage{textcomp}
\usepackage{xcolor}
\usepackage{siunitx}
\usepackage{textcomp}
\usepackage{hyperref}

\usepackage[justification=centering]{caption}

\begin{document}

\title{\LARGE \bf
Analyzing the Impact of Class Transitions on the Design of Pattern Recognition-based Myoelectric Control Schemes
\thanks{*This work was supported by the Natural Sciences and Engineering Research Council of Canada (DG 2014-04920).}
\thanks{$^{1}$All authors are with the Department of Electrical and Computer Engineering, University of New Brunswick, Canada,
        {\tt\small stallam@unb.ca, dmac@unb.ca, escheme@unb.ca}}%
\thanks{$^{2}$E. Scheme is also with the Institute of Biomedical Engineering, University of New Brunswick, Canada,
        {\tt\small escheme@unb.ca}}%
}

\author{Shriram Tallam Puranam Raghu$^{1}$, Dawn T. MacIsaac$^{1}$, 
and Erik J. Scheme$^{2}$,\IEEEmembership{~Senior~Member,~IEEE}}

\maketitle

\begin{abstract}
Despite continued efforts to improve classification accuracy, it has been reported that offline accuracy is a poor indicator of the usability of pattern recognition-based myoelectric control. One potential source of this disparity is the existence of transitions between contraction classes that happen during regular use and are reported to be problematic for pattern recognition systems. Nevertheless, these transitions are often ignored or undefined during both the training and testing processes. In this work, we propose a set of metrics for analyzing the transitions that occur during the voluntary changes between contraction classes during continuous control. These metrics quantify the common types of errors that occur during transitions and compare them to existing metrics that apply only to the steady-state portions of the data. We then use these metrics to analyze transition characteristics of 6 commonly used classifiers on a novel dataset that includes continuous transitions between all combinations of seven different contraction classes. Results show that a linear discriminant classifier consistently outperforms other conventional classifiers during both transitions and steady-state conditions, despite having an almost identical offline performance. Results also show that, although offline training metrics correlate with steady-state performance, they do not correlate with transition performance. These insights suggest that the proposed set of metrics could provide a shift in perspective on the way pattern recognition systems are evaluated and provide a more representative picture of a classifier’s performance, potentially narrowing the gap between offline performance and online usability.
\end{abstract}

Keywords - Surface Electromyography (SEMG); SEMG pattern recognition; evaluation metrics; transition metrics; usability; offline accuracy.

\section{Introduction}

Pattern recognition (PR) of surface electromyography (SEMG) signals is an active research area in several domains due to the wealth of information it provides. One of the most common applications of SEMG-PR is myoelectric control, such as is used to control prosthetic devices \cite{scheme_electromyogram_2011, amsuss_self-correcting_2014}. Compared to conventional EMG control of prostheses, which typically uses only the amplitude information from one or two SEMG signals, PR-based systems leverage more information from multiple signals by extracting temporal and spatial features, enabling the discrimination of more movements \cite{englehart_robust_2003}. SEMG-PR has also been explored for use in rehabilitation \cite{khokhar_surface_2010}, and more recently, human-computer interaction (HCI) based applications such as gesture recognition  \cite{chen_hand_2017} and computer game interaction \cite{nacke_biofeedback_2011}.

Despite the advantages of SEMG-PR, several factors have been shown to significantly affect the performance of SEMG-PR systems \cite{scheme_robustness_2014, vidovic_improving_2016}. These include confounding influences such as limb position, electrode shift, fatigue, changes in patterns over time, and in general, noisy decision streams resulting from the inherently stochastic EMG source  \cite{campbell_current_2020}. Many groups have endeavored to improve the robustness of SEMG-PR systems using a variety of approaches. Some have designed robust feature sets \cite{tkach_study_2010, hudgins_new_1993}, whereas others have employed filtering techniques to improve the quality of the underlying SEMG signal  \cite{gradolewski_arm_2015}. Post-processing techniques such as majority vote (MV) \cite{englehart_robust_2003} or rejection \cite{scheme_confidence-based_2013} have also been explored to improve the quality of the decision stream, particularly during periods of transition. Recently, temporal deep learning algorithms such as Recurrent Neural Networks (RNNs) and Temporal Convolutional Networks (TCNs) have achieved high classification accuracies by exploiting the temporal information in the EMG signals \cite{hu_novel_2018, betthauser_stable_2019}. These algorithms have been shown to outperform traditional classifiers as well as more spatial deep learning algorithms such as convolutional neural networks (CNNs) and thus, are being actively researched for use in SEMG-PR systems.

The relatively high accuracies reported in these and other SEMG-PR studies are encouraging, but research has also shown that offline accuracy does not generally predict online usability \cite{kyranou_causes_2018, hargrove_real-time_2007}. Consequently, many researchers now employ usability tests such as a Fitts law test  \cite{hargrove_control_2018, scheme_validation_2013} to test the benefits of proposed techniques. While it is of great importance to measure performance during online usage, the value in these tests is limited to doing just that. They offer little insight into the apparent disconnect with offline accuracy or any possible mechanisms for observed improvements \cite{robertson_effects_2019}. Because the user is ‘in-the-loop’ (responding to controller-specific feedback), they also preclude the subsequent use of collected data to compare or evaluate other control schemes.  
 
Some studies have pointed to the ambiguous transition period between contractions, in which no pre-trained class may be represented, as a possible explanation for disagreement between offline accuracy and online usability \cite{simon_decision-based_2011, scheme_training_2013}. Indeed, most offline studies collect static and/or separate sets of motions as training and testing data, from which classifiers are built and evaluated. However, the practical use of SEMG-PR control systems requires continuous and dynamic transitions from one contraction to another. Training and evaluating classifiers using disjointed contractions may, therefore, not capture information that explains or accounts for the dynamics seen during usability.

Hudgins et al., in their seminal paper \cite{hudgins_new_1993}, acknowledged the involvement of dynamics by training with ramp contractions, and others have since followed \cite{hargrove_real-time_2007, englehart_classification_1999}. Increased usability has been observed when training the classifier with ramp contractions as opposed to static contractions, despite a corresponding decrease in offline classifier accuracy. Scheme et al. explicitly evaluated the impact of contraction dynamics, and their results emphasized the importance of incorporating dynamics in the training process  \cite{scheme_training_2013}.

Although this has led to notable improvements, ramp contractions only model a subset of the dynamics - transitioning from rest to active contraction within a single motion class. Despite the prevalence of errors during transitions between motions \cite{robertson_effects_2019}, few studies have explicitly included or analyzed transitions from one motion class to another as part of these dynamics \cite{cote-allard_deep_2019}. Consequently, current design recommendations may be based on an overemphasis on steady-state classification. Incorporating all transitions during the training process may not be clinically practical given the large number of possible transition combinations (from each type of contraction to every other type), but an understanding of how these transitions impact current control schemes is warranted. Studying these transitions, however, is not straightforward.

When training and testing with static contractions to represent a motion class, segmentation of data is often trivialized by the collection process (stationary contractions are held for long periods and separated by a period in a neutral position). However, the segmentation, labeling, and evaluation of continuous data that include transitions pose a challenge, as it is not always easy to infer the start or endpoints of the motion classes \cite{asghari_oskoei_myoelectric_2007}. While inertial measurement units \cite{wolf_decoding_2013}, motion gloves, and motion capture systems \cite{pradhan_integration_2007} may be used to capture this intent, each approach has its own set of limitations and added complexity when trying to synchronize and align with corresponding EMG data \cite{ameri_real-time_2014}. Furthermore, differences in the relationship between the kinetics and kinematics observed by these systems during eccentric and concentric contractions further complicate the timing relationship with EMG \cite{ewins_clinical_2014}. In part because of these challenges, transitions have typically been ignored during training and offline evaluation of SEMG-PR.

A handful of studies have explicitly considered the temporal nature of the decision stream in SEMG-PR systems. For instance, Simon et al. \cite{simon_decision-based_2011} used a decision-based velocity ramp to modulate the proportional control output to mitigate the effects of transitions. Betthauser et al. \cite{betthauser_stable_2019} found TCNs to provide a more stable decision stream as compared to conventional classifiers when dealing with continuous datasets. While these and other studies have successfully leveraged continuous data, they have not comprehensively analyzed the transitions between classes. Moreover, there is currently no established framework that enables the analysis of continuous EMG transition data in an offline and repeatable way. Gaining a better understanding of performance during transitions may enable researchers to design SEMG-PR systems that better deal with transitions. However, due to the absence of such frameworks, the use of continuous data has been limited. As a result, transitions continue to be problematic during online control \cite{scheme_comparison_2015}, and their influence on usability remains unclear. 
As a step towards addressing this gap, in this work, we propose a methodology to collect a continuous dataset and a set of new metrics for evaluating the continuous dataset offline. This shift in perspective on how SEMG-PR systems are evaluated may reveal improvements not captured by conventional evaluation data and metrics. This may ultimately lead to the design of more robust pattern recognition-based myoelectric control. Our contribution to the field is therefore as follows:

\begin{itemize}
    \item We propose a methodology for collecting continuous test datasets for offline analysis that focuses on capturing both the steady-state as well as transitions between motion classes.
    \item We propose a set of metrics with which to analyze the continuous test dataset. This includes traditional accuracy metrics for assessing performance in steady-state as well as new metrics for assessing performance in transitions. Our intention with these metrics is to gain deeper insight into the characteristics of transitions and to facilitate a more comprehensive comparison of different algorithms using the same dataset.
    \item As a proof-of-concept, we compare the performance of 6 different classifiers using the same training and continuous test dataset. The results indicate clear differences across the classifiers which were not readily apparent when looking at only the traditional metrics, thus demonstrating the usefulness of the proposed methodology and metrics. 

\end{itemize}

\section{Methods}

\subsection{Data Acquisition}

Data were collected from ten able-bodied participants recruited from a graduate student population. All participants gave informed consent as approved by the University of New Brunswick’s Research Ethics Board (REB \#2015-134).

Eight bipolar channels of SEMG signals were recorded using the custom UNB smart-electrode cuff \cite{wilson_bus-based_2011}. Data were sampled at $\SI{1}{\kilo\hertz}$ with a $12$-bit resolution, and electrode pairs were placed equidistantly around the participant’s dominant forearm, proximal to the elbow. The signals were recorded on a desktop computer running a MATLAB\textregistered based data collection program \cite{scheme_flexible_2008}. Signals were digitally filtered to remove any power-line interference using a $3$rd order Butterworth band-stop filter at \SI{60}{\hertz}, \SI{180}{\hertz}, and \SI{300}{\hertz}.

\subsection{Training and Testing Protocol}

Four sets of conventional training repetitions were recorded from each participant. Each set comprised a repetition of each contraction class, beginning from a neutral position and ramping up to a moderate level of contraction. Three-second repetitions of seven classes were included: No Movement (NM), Wrist Flexion (WF), Wrist Extension (WE), Wrist Pronation (WP), Wrist Supination (WS), Chuck Grip (CG), and Hand Open (HO). Participants were prompted by a static image on a computer monitor to initiate each contraction and were guided for pace by a $3$-second countdown timer. An additional $3$-seconds were given at the end of each contraction to give participants time to return to a neutral position, rest, and prepare for the next contraction. Data were collected only during the prompted contractions and used to train the classifiers outlined below. This resulted in a training data size of $10$ participants x $7$ contraction classes/participant x $4$ repetitions/contraction class x $3$ seconds/repetition, for a total of $840$ seconds of training data. 

\begin{figure*}[htb]
    \centering
    \includegraphics[width=\textwidth,height=\textheight,keepaspectratio]{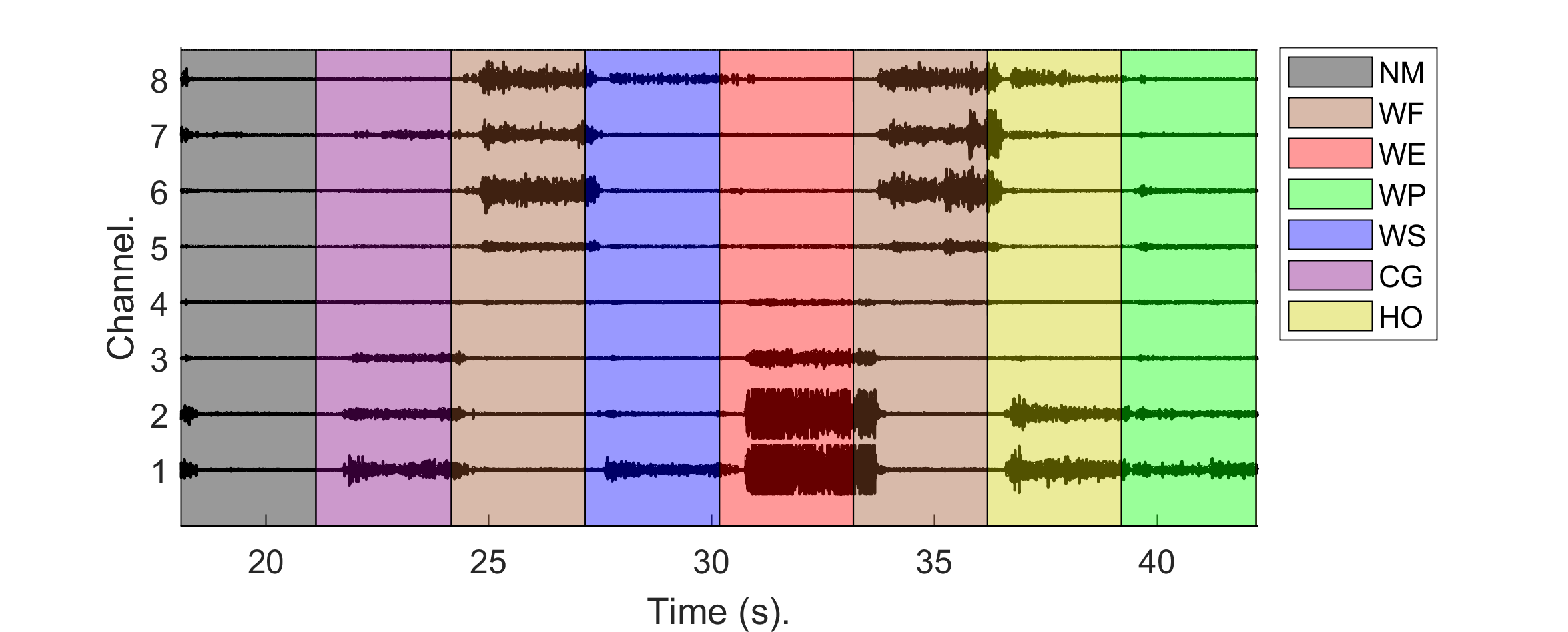}
    \caption{A $24$-second SEMG time series segment obtained from a test set. The shaded regions indicate the prompted contraction}
    \label{fig:Raw_SEMG}
\end{figure*}

Three sets of continuous transition test data were also collected. In these test sets, participants were prompted using the same visualizations as before, but with no delay between motions. This required them to directly transition from one class to another without intentionally returning to a neutral position between transitions, and without any knowledge of the next class. When the participants completed a transition from one class to the next, they were instructed to maintain that contraction (in steady-state) until the next class was prompted. At that time, they were asked to change to the new class immediately. Each set began in the no-motion class and included all combinations of transitions from each class to every other class. Prompts were changed every $3$ seconds and, other than ensuring that all combinations were included, transitions were randomly ordered across sets and participants so that participants could not anticipate the next class. No feedback was provided to the participants, and they were encouraged to transition to each new prompt as quickly as comfortably possible. The SEMG signals were recorded continuously throughout each full set. This resulted in a test data size of $10$ subjects x $3$ trials/subject x $42$ prompts/trial x $3$ seconds/prompt, for a total of $3780$ seconds of test data. The $42$ prompts in each test trial are a result of transitions from all $7$ contraction classes to all other classes, and therefore resulting in a total of $7 \times 6 = 42$ transitions. A section of SEMG signals obtained from a set of representative contractions is shown in Figure \ref{fig:Raw_SEMG}, with the prompted contraction and timing shaded in different colours

All signals were segmented with a frame length of \SI{160}{\milli\second} and an increment of \SI{16}{\milli\second} for classification. Hudgins time-domain feature set \cite{hudgins_new_1993} (Mean Absolute Value (MAV), Slope Sign Changes (SSC), Waveform Length (WL) and Zero Crossings (ZC)) was computed for each frame per channel, resulting in a total of $32$ features ($4$ features * $8$ channels) per frame. 

\subsection{Comparing Classifier Performance}
\label{sec:comp_classifier_perf}
The performance of several commonly used classification techniques was compared in this work. These included Linear Discriminant Analysis (LDA), Quadratic Discriminant Analysis (QDA), Support Vector Machine with Linear Kernel (SVM), K-Nearest Neighbours (KNN), Multilayer Perceptron-Artificial Neural Network (MLP-ANN) and Random Forest (RF) \cite{scheme_electromyogram_2011, zhang_real-time_2019}.These represent a range of popular generative and discriminative classifiers, and they have also been shown to have excellent offline performance in SEMG-PR \cite{scheme_electromyogram_2011, oskoei_support_2008, phinyomark_emg_2013} and were therefore chosen for this proof-of-concept study.

LDA was implemented as described in \cite{scheme_confidence-based_2013}, SVM was implemented using LIBSVM (version 3.23), MLP-ANN was implemented using MATLAB\textregistered Deep Learning Toolbox, and the rest of the classifiers were implemented using the MATLAB\textregistered Statistics and Machine Learning Toolbox. The MATLAB\textregistered default settings were used in the implementation of the MLP-ANN and RF. MATLAB\textregistered 2019a was used to conduct the analysis.


    
        
        
        
        

The hyperparameters for the various classifiers were chosen based on 10-fold cross-validation using the offline training data of two randomly chosen participants. For MLP-ANN, we gradually increased the size of the hidden layer, and for hidden layer size greater than $5$ neurons, we found marginal performance improvements ($<1\%$), and therefore the hidden layer size was set to $5$. For KNN, SVM, and RF, a range of values were found to provide comparable performance, and so for SVM and RF, we used their default hyperparameter values ($c =1$ and Learning Cycles of $100$ respectively) which were in the range of values that resulted in high performance. Finally, for KNN, we set $k = 5$ based on suggestions by Kim et al in \cite{kim_comparison_2011}, which was consistent with the range of values identified in our pilot work. 

An essential aspect of supervised machine learning is the creation of labels to train and evaluate classifier performance. In continuous transition data, however, this process is not trivial. Labels can readily be assigned to the steady-state portion of signals, but the labeling of frames within the dynamic portions of contraction may depend on the previous class, the reaction time of the user, the speed of transition, and the kinematics of the contraction. Additionally, because multi-degree of freedom joint control is continuous, the use of discrete classes may lead to undefined periods during these transitions \cite{wurth_real-time_2014}. For these reasons, in this work, the transition and steady-state regions were analyzed separately. 

To identify the steady-state regions, which most classifiers have been shown to accurately classify \cite{hargrove_real-time_2007}, each classifier’s decision stream was smoothed by taking a $9$-frame majority vote across the stream. The beginning of steady-state was then defined as the first frame after a prompt whose classification decision corresponded to the prompted class. In the rare instance when no corresponding frames were identified during a prompt, it was assumed that an operator error had occurred (e.g., the participant elicited the wrong contraction), and the contraction was discarded from the analysis. The same process was followed to find the end of steady-state, which always occurred shortly after the prompt for the upcoming contraction. The end frame for each steady-state period was defined as the first frame after a prompt whose classification decision did not correspond to the previous prompt class. Finally, to compensate for the delay introduced by the majority vote, all steady-state labels were adjusted backward by $4$ frames. 

The start and end frames of the steady-states were then used to label the frames. All frames that fell between a given start and endpoint were assigned the label corresponding to the prompt. All frames that fell between consecutive steady-states were then defined as transitions and were not assigned labels. Because these frames had no labels, traditional accuracy metrics could not be applied to transition regions, and so a new set of performance metrics were developed. The metrics for assessing classifier performance during both steady-state and transition are described in the next section.

\subsection{Assessing the performance of the classifiers}
Leave-one-out cross-validation accuracy, commonly used to assess offline classifier performance in SEMG-PR studies \cite{zia_ur_rehman_multiday_2018}, was used to perform a participant-wise leave-one-set-out analysis of the training data. This was done to establish a baseline that is comparable with previous literature and to confirm that the classifiers were properly trained. In each fold, a classifier was trained using all but one training set and tested with the remaining set. This process was repeated until all the sets were assessed, and the average error across all sets and participants was recorded as the offline training error. This process was completed for each of the six classifiers and the results were analyzed for significant differences with a Kruskal-Wallis test as the distributions of offline training errors failed normality testing.
Next, the test sets were used to evaluate each classifier’s steady-state and transition performance. The proposed metrics facilitate the comparison of classifier performance across the same set of steady-state contractions and transitions. Because each classifier is evaluated using the same data, unlike conventional usability tests (where each system must be tested separately due to the influence of user-in-the-loop control), results can be compared between classifiers, alleviating the need for a deterministic ground truth. That is to say, there is value in knowing which classifiers respond more quickly to changes and with less volatility than others.

\subsubsection{Steady-state metrics}

Commonly used metrics were used to assess the steady-state regions as described below \cite{campbell_linear_2019}. 


\renewcommand{\theenumi}{\alph{enumi}}

\begin{enumerate}
 
    \item \textbf{Total Error Rate (TER)}: the percentage of incorrect decisions in a steady-state region.  Lower values are better.
    
	\item \textbf{Active Error Rate (AER)}: The percentage of decisions corresponding to an active (i.e. any class besides no motion) but incorrect class. Lower values are better.
	
	\item \textbf{Instability (INS)}: the number of times, in a steady-state region, that a pair of consecutive active class decisions differs, normalized by the total number of decisions in that region. Lower values are better.
	
\end{enumerate}

\subsubsection{Transition metrics}

In total, six transition metrics were employed to measure the transition performance as described below. The first three metrics focus on the response times of the classifiers. The last three focus on fluctuations within transition regions. A common phenomenon during transitions that degrades usability is the rapid and stochastic fluctuation between classifier decisions \cite{robertson_effects_2019}. As in the steady-state metrics, instability captures these rapid changes, and two related metrics, tertiary class error, and percent no movement, help to further discern how these changes manifest.

\begin{enumerate}

    \item \textbf{Offset delay ($\boldsymbol{T_{OFFSET}}$)}: the delay between a change in prompt and when the classifier leaves the previous steady state. Lower values are better.

    \item \textbf{Onset delay ($\boldsymbol{T_{ONSET}}$)}: the delay between a change in prompt and when a classifier reaches the corresponding steady-state. Lower values are better.
    
	\item \textbf{Transition Duration ($\boldsymbol{T_{TRANSITION}}$)}: the duration between two consecutive steady states. Lower values are better.
	
	\item \textbf{Instability (INS)}: the number of times, in a region of transition, that a pair of consecutive active class decisions differ, normalized by the total number of decisions in that region. Lower values are better.
	
	\item \textbf{Tertiary Class Error (TCE)}: the number of decisions in a transition region that do not correspond to the steady states adjacent to the region, or to NM, normalized by the total number of decisions in that region. TCE can be considered similar to AER, but for transitions. Lower values are better.
	
	\item \textbf{Percent No Movement (PNM)}: the number of NM decisions in a transition region, normalized by the total number of decisions in that region. Higher values are better. 
\end{enumerate}

\begin{figure*}[htb]
    \centering
    \includegraphics[width=\textwidth,height=\textheight,keepaspectratio]{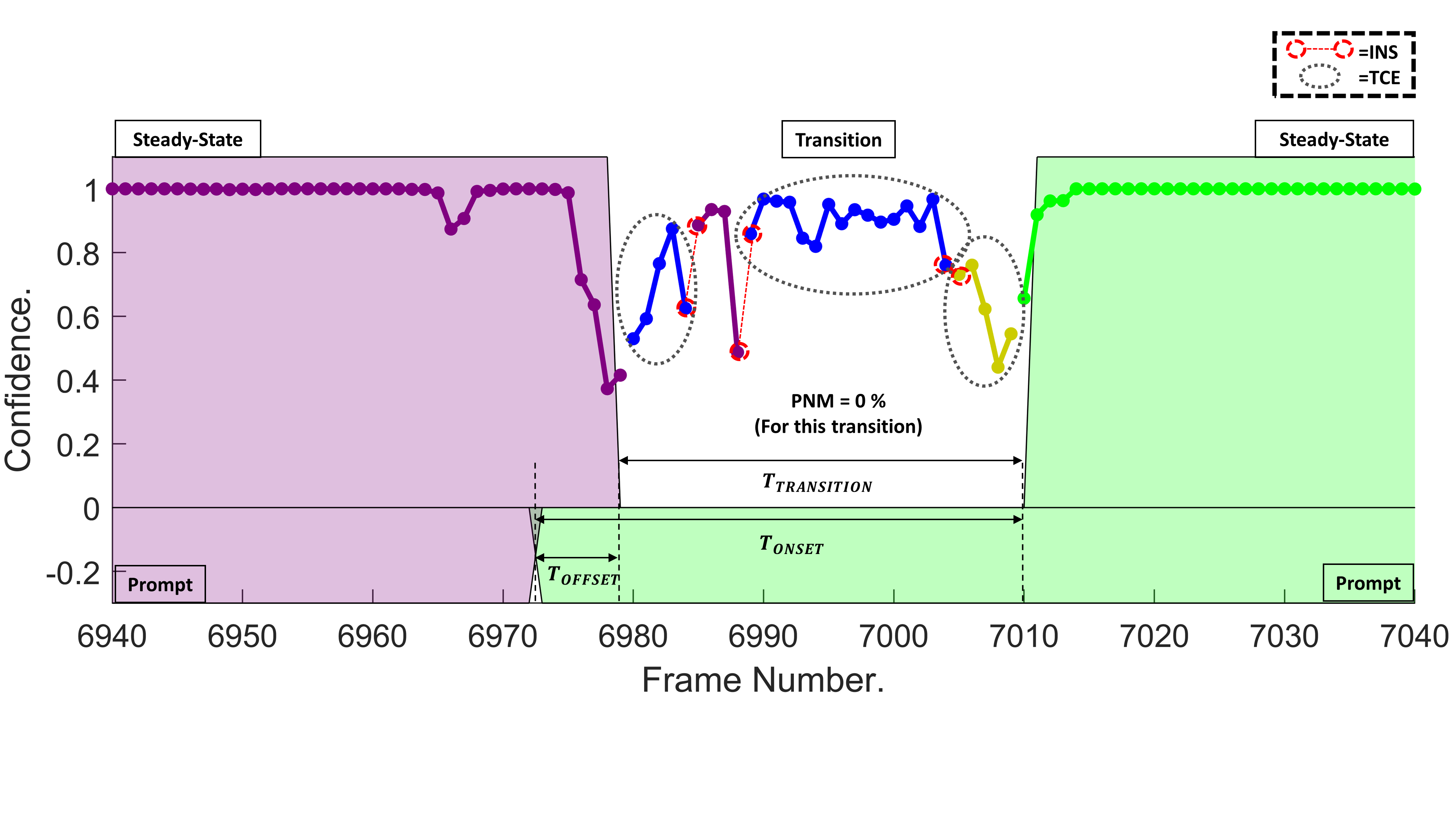}
    \caption{Visual depiction of transition metrics for an example decision stream.}
    \label{fig:Metrics_Visual}
\end{figure*}

Figure \ref{fig:Metrics_Visual} depicts the transition metrics for an example decision stream. The figure shows a typical transition from one active class to another. In the figure, each data point represents a colour-coded confidence in the classification decision for a frame. The colour of the data point reflects the class decision. The lower shaded regions indicate the prompted classes by colour (with a change in the prompt at frame number $6973$). The upper shaded regions indicate the assigned labels by colour (obtained as explained in Section \ref{sec:comp_classifier_perf}). When the colour of the data points matches the colour of the upper shaded regions, the classification decision matched the frame label. Data in the example figure were produced using an LDA classifier, but the figure is representative of all classifiers. 

These metrics were computed and averaged across the three continuous test trial sets on a per participant and per transition basis. The transitions were then grouped into three categories: transitions from an active class to rest (A2R), transitions from rest to an active class (R2A), and transitions from one active class to another (A2A).  This process was completed for each of the six classifiers and the results were analyzed for significant differences between groups and classifiers using a Kruskal-Wallis test. When significant differences were indicated, a post-hoc pairwise multiple comparison test with Dunn-Šidák correction was applied.  Alpha values were always set to $0.05$. 

Finally, a Pearson’s correlation analysis was performed to ascertain if the offline TER correlated to any of the metrics used to evaluate the test data.

\section{Results}

The training errors are provided for the different classifiers in Table \ref{tbl:Offline_Error}. Results show that all classifiers performed comparably in terms of TER on the training set. RF obtained the lowest TER of $12.7\%$, but no significant differences were found ($p > 0.9$). The errors obtained were also within the reported ranges in the wider literature \cite{scheme_training_2013}, suggesting that the classifiers were trained appropriately.

\begin{table}[htbp]
    \begin{center}
    \begin{tabular}{c|c}
    
        \hline
        
        \textbf{Classifier}  & \textbf{Training TER (\%)} \\
        
        \hline
        
        ANN	                &   $15.1 \pm 8.6$ \\
        KNN	                & $ 15.8 \pm 8$  \\
        LDA	                & $14.1 \pm 7.3$ \\
        QDA	                & $14.1 \pm 7.8$ \\
        RF	                & $12.7 \pm 6.5$ \\
        SVM	                & $14.5 \pm 8.1$ \\
        
        \hline
    \end{tabular}
    \end{center}
    \caption{\label{tbl:Offline_Error} Comparison of offline training error (mean and standard deviation) (\%).}
\end{table}

Table \ref{tbl:Steady_State_Error} shows the steady-state metrics obtained from the continuous test set. Unlike in training, these results show that LDA was the best classifier across all three metrics, while QDA had the worst AER and TER, and KNN had the worst INS. Despite the offline training error being almost identical for LDA and QDA, the differences in the steady-state metrics were substantive. Statistical differences were found across classifiers for AER and INS ($p < 0.05$) but not TER ($p > 0.9$). The post-hoc analysis indicated that LDA had significantly lower INS than all other classifiers and lower AER than QDA and KNN ($p < 0.05$ in both cases). No other significant differences were noted.

\begin{table}[tb]
    \begin{center}
    \begin{tabular}{c|ccc}
    
        \hline
        
        \textbf{Classifier}  &   \textbf{AER (\%)}              &   \textbf{TER (\%)}                                     &   \textbf{INS (\%)}  \\
        
        \hline
        
        ANN         &   $14.7 \pm 10.3$                         &   $19.3 \pm 13.4$                              &   $3.0 \pm 2.2$   \\
        
        KNN	        &   $16.2 \pm 11.1$                         &   $20.1 \pm 13.2$                              &   $3.9 \pm 2.9$   \\
        
        LDA         &   $\boldsymbol{11.2 \pm 9.3} {}^{+}$      &   $\boldsymbol{18.7 \pm 13.8}$                 &   $\boldsymbol{2.0 \pm 1.9{}^{++}}$   \\
        
        QDA	        &   $18.1 \pm 14.2$                         &   $21.3 \pm 15.7$                              &   $3.3 \pm 2.9$   \\
        
        RF          &   $15.4 \pm 11.9$                         &   $20.4 \pm 14.7$                              &   $3.5 \pm 2.7$   \\
        
        SVM         &   $14.2 \pm 10.5$                         &   $19.2 \pm 14.0$                              &   $3.0 \pm 2.4$   \\
        \hline
    \end{tabular}
    \end{center}
    
    \caption{\label{tbl:Steady_State_Error}Comparison of steady state metrics for different classifiers (mean and standard deviation). Bolded values indicate best performers. ++Indicates significantly better than all other classifiers; +indicates significantly better than KNN and QDA.}
    
\end{table}

When analyzing the transition metrics, first, a preliminary analysis was done to compare the metrics from the three transition groups, pooling data from all the classifiers and participants. The results indicated that $T_{TRANSITION}$, INS, and TCE were smaller for the R2A case compared to the other cases. The A2R case was second-best, with slightly lower  $T_{TRANSITION}$ and INS, and higher PNM, compared to the A2A group. A Kruskal-Wallis test yielded statistically significant differences across the transition groups ($p < 0.001$) and the post-hoc pairwise multiple comparison tests confirmed the observation that R2A outperformed both other groups and A2R outperformed A2A ($p < 0.05$). These results indicate that transitions from rest to active classes are handled best by the classifiers, followed by transitions from active to rest. These results were also found to be consistent across each type of classifier.

\begin{table*}[htbp]
    \begin{center}
    \begin{tabular*}{\textwidth}{c|@{\extracolsep{\fill}}cccccc}
    
    \hline
    
    \textbf{Classifier}	                            &   $\boldsymbol{T_{OFFSET}}$\textbf{(ms)}           &   $\boldsymbol{T_{ONSET}}$\textbf{(ms)}   &   $\boldsymbol{T_{TRANSITION}}$\textbf{(ms)}       &   \textbf{INS (\%)}                       &   \textbf{TCE (\%)}                                &   \textbf{PNM (\%)}    \\
    
    \hline
    
    ANN     &   $444.2 \pm 124.3$	                    &   $\boldsymbol{504.2 \pm 138.0}$                   &   $60.0 \pm 41.6$                         &   $2.9 \pm 4.8$                                    &	$10.0 \pm 13.6$                         &	$0.3 \pm 1.4$   \\
    
    KNN     &   $\boldsymbol{441.7 \pm 127.4}$          &   $527.5 \pm 169.8$                                &   $82.2 \pm 74.6$                         &   $6.6 \pm 8.5$                                    &   $17.8 \pm 20.0$                         &	$0.4 \pm 2.3$   \\
    
    LDA     &   $507.6 \pm 138.8$                       &   $561.0 \pm 144.6$                                &   $\boldsymbol{53.4 \pm 36.2} {}^{+}$     &   $\boldsymbol{2.0 \pm 3.7} {}^{+}$                &   $\boldsymbol{7.3 \pm 12.1}{}^{+}$       &   $0.3 \pm 2.1$   \\
    
    QDA     &   $444.1 \pm 156.7$                       &   $504.7 \pm 169.7$                                &   $108.9 \pm 108.8$                       &   $5.0 \pm 6.6$                                    &   $21.4 \pm 24.1$                         &   $0.3 \pm 2.2$   \\
    
    RF      &   $450.8 \pm 135.6$                       &   $534.7 \pm 204.5$                                &   $84.0 \pm 116.0$                        &   $3.3 \pm 6.8$                                    &   $12.3 \pm 19.3$                         &   $\boldsymbol{1.0 \pm 5.0}$  \\
    
    SVM     &   $451.3 \pm 125.9$                       &   $518.1 \pm 135.2$                                &   $66.8 \pm 52.9$                         &   $3.9 \pm 6.9$                                    &   $11.3 \pm 16.8$                         &   $0.6 \pm 2.8$   \\
    
    \hline
    \end{tabular*}
    \end{center}
    
    \caption{\label{tbl:R2A_Transition_Error}Comparison of transition metrics across the different classifiers in the R2A transition group (mean and standard deviation). + indicates significantly better than at least one classifier, - indicates worse than at least one classifier.}
    
\end{table*}

\begin{table*}[htbp]
    \begin{center}
    \begin{tabular*}{\textwidth}{c|@{\extracolsep{\fill}}cccccc}
    
    \hline
    
    \textbf{Classifier}	                            &   $\boldsymbol{T_{OFFSET}}$\textbf{(ms)}           &   $\boldsymbol{T_{ONSET}}$\textbf{(ms)}   &   $\boldsymbol{T_{TRANSITION}}$\textbf{(ms)}       &   \textbf{INS (\%)}                       &   \textbf{TCE (\%)}                                &   \textbf{PNM (\%)}    \\
    
    \hline
    
    ANN         &   $221.6 \pm 195.6$                   &	$456.3 \pm 293.6$                                    &   $250.1 \pm 237.3$                   &   $4.3 \pm 5.3$                                        &   $42.3 \pm 31.8$	                    &   $55.7 \pm 31.5$ \\
    
    KNN	        &   $225.4 \pm 216.5$                   &   $453.2 \pm 252.1$                                    &   $245.6 \pm 229.3$                   &   $6.1 \pm 6.8$                                        &   $42.0 \pm 29.0$                     &   $55.2 \pm 29.4$ \\
    
    LDA         &   $\boldsymbol{185.9 \pm 173.8}$      &   $\boldsymbol{230.4 \pm 236.5}{}^{++}$                &   $\boldsymbol{87.1 \pm 96.4}{}^{++}$ &   $\boldsymbol{1.5 \pm 3.8}{}^{++}$                    &   $\boldsymbol{15.5 \pm 21.8}{}^{++}$ &   $\boldsymbol{83.8 \pm 22.5}{}^{++}$ \\
    
    QDA         &   $279.8 \pm 204.9$                   &   $649.3 \pm 283.2$                                    &   $409.4 \pm 307.4$                   &   $6.0 \pm 5.2$                                        &   $56.4 \pm 29.0$                     &   $39.2 \pm 26.5$ \\
    
    RF          &   $202.4 \pm 221.5$                   &   $438.3 \pm 243.1$                                    &   $252.9 \pm 254.0$                   &   $5.5 \pm 6.4$                                        &   $41.7 \pm 29.3$                     &   $56.5 \pm 29.9$ \\
    
    SVM         &   $210.5 \pm 199.6$                   &   $437.3 \pm 274.3$                                    &   $238.4 \pm 205.3$                   &   $4.1 \pm 4.7$                                        &   $42.0 \pm 29.0$                     &   $55.3 \pm 29.5$ \\
    
    \hline
    \end{tabular*}
    \end{center}
    
    \caption{\label{tbl:A2R_Transition_Error}Comparison of transition metrics across the different classifiers in the A2R transition group (mean and standard deviation). Bolded values indicate best performers. ++Indicates significantly better than all other classifiers.}
    
\end{table*}

\begin{table*}[htbp]
    \begin{center}
    \begin{tabular*}{\textwidth}{c|@{\extracolsep{\fill}}cccccc}
    
    \hline
    
    \textbf{Classifier}	                            &   $\boldsymbol{T_{OFFSET}}$\textbf{(ms)}           &   $\boldsymbol{T_{ONSET}}$\textbf{(ms)}   &   $\boldsymbol{T_{TRANSITION}}$\textbf{(ms)}       &   \textbf{INS (\%)}                       &   \textbf{TCE (\%)}                                &   \textbf{PNM (\%)}    \\
    
    \hline
    
    ANN         &   $163.2 \pm 318.5$                   &   $491.8 \pm 217.2$                                    &   $331.9 \pm 318.1$                   &   $7.9 \pm 6.5$                                        &   $32.6 \pm 25.3$                     &   $18.9 \pm 24.0$ \\
    
    KNN	        &   $175.0 \pm 284.3$                   &   $\boldsymbol{479.9 \pm 221.5}$                       &   $\boldsymbol{325.5 \pm 310.5}$      &   $11.1 \pm 8.2$                                       &   $35.9 \pm 26.6$                     &   $15.0 \pm 20.1$ \\
    
    LDA         &   $\boldsymbol{150.6 \pm 284.1}$      &   $533.7 \pm 198.4 {}^{--}$                            &   $404.5 \pm 296.6{}^{--}$            &   $\boldsymbol{5.0 \pm 5.4}{}^{++}$                    &   $\boldsymbol{19.9 \pm 21.2}{}^{++}$ &   $\boldsymbol{42.1 \pm 26.4}{}^{++}$ \\
    
    QDA         &   $221.0 \pm 305.4$                   &   $523.7 \pm 297.5 {}^{*}$                             &   $337.2 \pm 334.6$                   &   $9.3 \pm 7.5$                                        &   $39.1 \pm 28.2$                     &   $7.6 \pm 15.5{}^{--}$  \\
    
    RF          &   $184.0 \pm 306.0$                   &   $492.1 \pm 240.4$                                    &   $336.1 \pm 330.6$                   &   $9.6 \pm 8.7$                                        &   $33.3 \pm 26.2$                     &   $18.0 \pm 22.8$ \\
    
    SVM         &   $174.0 \pm 322.5$                   &   $495.6 \pm 226.8$	                                 &   $339.7 \pm 322.2$                   &	$7.4 \pm 6.4$                                         &   $32.2 \pm 25.9$                     &   $19.7 \pm 24.8$ \\
    
    \hline
    
    \end{tabular*}
    \end{center}
    
    \caption{\label{tbl:A2A_Transition_Error}Comparison of transition metrics across the different classifiers in the A2A transition group (mean and standard deviation). Bolded values indicate best performers. ++Indicates significantly better than all other classifiers, -- indicates significantly worse than all other classifiers, and * indicates excluded from the comparison.}
    
\end{table*}

Next, for each of the transition groups, a comparison of classifier performance was conducted. The results of each of these comparisons are shown in Tables \ref{tbl:R2A_Transition_Error}, \ref{tbl:A2R_Transition_Error} and \ref{tbl:A2A_Transition_Error}, for R2A, A2R, and A2A respectively. The results obtained show that no single classifier performed best across all metrics in all $3$ groups, but that LDA stood out in many regards. First, LDA yielded the best performance for all metrics in the A2R group. A Kruskal-Wallis test indicated that there were statistical differences between the classifiers for each metric ($p < 0.05$) except for $T_{OFFSET}$ ($p > 0.1$). For those metrics yielding significant differences, post-hoc pairwise multiple comparison tests indicated that LDA significantly outperformed all other classifiers ($p < 0.05$).

In the R2A group, a Kruskal-Wallis test yielded statistical differences across the classifiers ($p < 0.05$) for all metrics except for $T_{ONSET}$ ($p > 0.25$) and PNM ($p > 0.8$).  For  $T_{TRANSITION}$, INS, and TCE, LDA again yielded the best results and post-hoc pairwise multiple comparison tests confirmed that the LDA outperformed QDA in $T_{TRANSITION}$, KNN in INS, and QDA and KNN in TCE ($p < 0.05$). However, QDA was significantly better than LDA in $T{OFFSET}$ ($p < 0.05$). No other classifier outperformed any other on more than one metric. 

Finally, in the A2A group, LDA was again the best performer in terms of   $T_{OFFSET}$, INS, TCE, and PNM. A Kruskal-Wallis test revealed significant differences for all of these metrics ($p < 0.05$) and the post hoc analysis revealed that LDA outperformed all other classifiers ($p < 0.05$). Significant differences were also found in  $T_{TRANSITION}$ and $T_{ONSET}$ for this transition group, and unlike the previous results, the post hoc analysis revealed that LDA performed worse compared to all other classifiers for $T_{TRANSITION}$, and all but QDA for $T_{ONSET}$ ($p < 0.05$).

It should be noted that using the timing of the prompts as the ground truth for computing delay estimates introduces unrealistically long biases (they include the user response delay). While the relative performance of classifiers remains unaffected and valid, the absolute numbers should be interpreted carefully with this understanding.

Finally, results from the correlation analysis for the steady-state and transition metrics are shown in Tables \ref{tbl:Corr_Steady_State} and \ref{tbl:Corr_Transition_Metrics} respectively. The offline TER showed a moderate, positive, and statistically significant correlation with all of the steady-state metrics ($p < 0.001$). Conversely, no consistent correlation was found across the transition metrics. Offline TER showed a moderate, positive, and statistically significant correlation with TCE and PNM in the R2A group ($p < 0.001$), but no other metrics showed any significant correlation for that group ($p > 0.1$). Offline TER showed no significant correlation with any of the transition metrics in the A2R group ($p > 0.35$) and, for the R2A group, only INS showed a statistically significant (albeit weak) correlation ($p < 0.05$ for INS, $p > 0.1$ for all the other metrics).

\begin{table}[htbp]
    \begin{center}
    \begin{tabular}{c|c}
        \hline
        
        \textbf{Metric}  &   \textbf{R} \\
        
        \hline
        
        AER     &   $0.58{}^{*}$    \\
        TER	    &   $0.56{}^{*}$    \\
        INS	    &   $0.54{}^{*}$    \\
        
        \hline
    \end{tabular}
    \end{center}
    \caption{\label{tbl:Corr_Steady_State}Correlation coefficient (R) between offline TER and steady-state metrics. * indicates statistically significant ($p < 0.05$) values.}
\end{table}

\begin{table}[htbp]
    \begin{center}
    \begin{tabular}{c|ccc}
    
        \hline
        
        \textbf{Metric}          &   $\boldsymbol{R_{R2A}}$   &   $\boldsymbol{R_{A2R}}$   &   $\boldsymbol{R_{A2A}}$   \\
        
        \hline
        
        $T_{OFFSET}$    &   $0.15$      &   $0.03$      &   $0.12$      \\
        $T_{ONSET}$	    &   $0.07$      &	$-0.12$     &	$0.19$      \\
        $T_{TRANSITION}$&	$0.09$      &	$-0.02$     &	$-0.05$     \\
        INS	            &   $0.21$      &	$0.09$      &	$0.26{}^{*}$\\
        TCE	            &   $0.55{}^{*}$&	$0.1$       &   $0.22$      \\
        PNM	            &   $0.46{}^{*}$&	$0.1$       &	$0.20$      \\
        
        \hline
    \end{tabular}
    \end{center}
    \caption{\label{tbl:Corr_Transition_Metrics}Correlation coefficient (R) between offline TER and transition metrics across the three groups. * indicates statistically significant ($p < 0.05$) values.}
\end{table}

\section{Discussion}

In this work, we highlighted the need to analyze transitions when evaluating classifiers and proposed a methodology of collecting continuous test data, as well as a set of metrics to evaluate the data. This simple approach can be used to quantify performance during transitions and steady-state regions of contractions separately to provide a more complete picture of the benefits of a particular classifier. We then used the metrics to compare 6 commonly used classifiers to determine how their performances differ between steady-state and transitions, and if a single classifier performs well across all metrics. We also compared these results to the commonly used leave-one-set-out offline training error of the classifiers.

The offline training error showed no significant differences across the classifiers, as has been shown in previous works \cite{hargrove_comparison_2007, adewuyi_evaluating_2016}. These results, however, did not translate to either of the steady-state or transition metrics obtained from the continuous transition test data. For instance, even though LDA and QDA had almost identical offline training performance, when performance was evaluated on the test data, LDA was frequently significantly better across most metrics (in both transition and steady-state), while QDA was often the worst. These discrepancies add to the argument that offline performance is an incomplete indicator of the overall performance of the PR system. 

When evaluating transition errors, there were significant differences across the three different groups of transitions (R2A, A2R, and A2A). Classifiers consistently performed better on the R2A group compared to the other two groups when considering $T_{TRANSITION}$, INS, and TCE. This is perhaps reflective of the inclusion of ramping contractions during training, which informs the classifiers of the dynamics of transition from rest to steady-state \cite{scheme_training_2013}. This group did yield a lower PNM compared to the other groups, which can sometimes indicate higher volatility, but only if this volatility is also supported by high INS and TCE. Since these metrics were also low for the R2A case, the low PNM may simply be a reflection of a low volatility transition.

The A2R transitions outperformed the A2A transitions in $T_{TRANSITION}$, INS, and TCE. Although the training data included only ramps from rest to active classes, this could still act to inform the classifiers about the boundaries between each class and rest. Conversely, the transitions between active contraction classes are not typically included during training because of the onerous number of necessary combinations (particularly for classifiers with many classes). These transitions may travel through the NM class (the first class is released before initiating the second) but may also involve traversing feature space through other class boundaries in undefined ways.

LDA is commonly used in the literature because it is simple to implement, fast to train, and reduces over-tuning \cite{negi_feature_2016}. Despite being touted for its simplicity, LDA was found to be the most promising classifier across all transition cases in this work. It had the lowest transition durations for the R2A and A2R groups, and, despite having the highest transition duration for the A2A group,  LDA had significantly lower INS and TCE, and significantly higher PNM, indicating a less volatile transition compared to the other classifiers. Likely, the generative and parametric nature of the LDA helps it generalize better than other classifiers when the test scenario includes more variability than seen during training. Although QDA is a similar generative classifier, it performs relatively poorly, even compared to the more discriminative classifiers. This may be a result of over-tuning because, unlike the LDA which pools covariance estimates, the QDA models each class separately. This supports that caution should be taken with more complex classifiers that may not generalize beyond training. 

The pooling assumption used by LDA, however, is not always advantageous as it may cause issues when classes have vastly different covariances. For instance, the NM class tends to have a much smaller covariance as compared to classes that involve active contractions. Using the pooled covariance effectively increases the time it takes the LDA to transition out of NM. This is evidenced by the relatively large $T_{OFFSET}$ in the R2A group. Fortunately, this effect is reversed when transitioning into NM, as supported by the small $T_{ONSET}$ in the A2R case. The overall strong performance of LDA suggests that accepting this trade-off may lead to better generalization. Further investigation into the transitions between different active classes indicated that all transitions are not necessarily equally classifiable. For instance, we found that transitions from any active class to wrist pronation resulted in double the TCE than those transitioning to wrist supination. We also noted that transitions to chuck grip were $5$ times faster than those to wrist extension. These vast differences in performance and behaviours during transitions warrant the continued exploration of temporal and state-based approaches for myoelectric control.

It is important to note that, in this study, the effect of different features was not explored. This is an important consideration, as features have been found to have a critical impact on overall classification performance \cite{hargrove_comparison_2007}. Here, we used the commonly cited Hudgins’ TD feature set to focus on classifiers, but many features have now been shown to perform better in various circumstances  \cite{phinyomark_navigating_2017}. Consequently, future work will explore how the suite of EMG features, either engineered or learning via deep learning, impacts transition behaviour. For this proof-of-concept study, data from $10$ participants were collected to construct the training and test datasets. Although the use of $10$ participants is relatively common in SEMG-PR research \cite{jun-uk_chu_real-time_2006, too_new_2018} the absolute performance of the classifiers may change with larger datasets. As such, for future works, we plan to expand the dataset by including more participants and also increase the size of the training and test data obtained from each participant. 

Finally, although the proposed metrics are capable of quantifying performance during transitions, it is not yet known if they will correlate with usability. As has been reported with online usability \cite{lock_real-time_2005}, in this work, we found no consistent correlation between offline accuracy and transition metrics across the transition groups. Specifically, the A2R and A2A groups showed weak to no correlation across all metrics, while two metrics in the R2A group (TCE and PNM) showed moderate correlation. While the lack of consistent correlation does not conclusively link transition performance and usability, it offers sufficient potential to warrant further investigation. We plan to fully study the correlation between usability and the transition metrics in upcoming studies.

The proof-of-concept study demonstrates that considering the performance during transitions, as measured by the proposed metrics, impacts the classifier choice, and therefore SEMG-PR design. Future work is being planned to expand this study to understand how all stages of the SEMG-PR pipeline, including pre-processing, feature extraction, and post-processing, are impacted when considering transitions, as well as the impact on online usability of SEMG-PR systems. 

\section{CRediT authorship contribution statement}
\textbf{Shriram Tallam Puranam Raghu}: Conceptualization, Methodology, Software, Investigation, Formal analysis, Data curation, Writing original draft, Writing - review \& editing, Visualization, Resources. \textbf{Dawn T. MacIsaac}: Conceptualization, Methodology, Software, Investigation, Formal analysis, Data curation, Writing – original draft, Writing - review \& editing, Visualization, Resources. \textbf{Erik J. Scheme}: Conceptualization, Methodology, Software, Investigation, Formal analysis, Data curation, Writing – original draft, Writing - review \& editing, Visualization, Resources, Funding acquisition.

\section{Declaration of Competing Interest}
The authors declare that they have no known competing financial interests or personal relationships that could have appeared to influence the work reported in this paper.

\section*{Acknowledgment}

We acknowledge the support of the Natural Sciences and Engineering Research Council of Canada (NSERC) [funding reference number DG 2014-04920].

\printbibliography
\end{document}